\documentclass{article}

\usepackage{arxiv}

\usepackage[utf8]{inputenc} % allow utf-8 input
\usepackage[T1]{fontenc}    % use 8-bit T1 fonts
\usepackage{hyperref}       % hyperlinks
\usepackage{url}            % simple URL typesetting
\usepackage{booktabs}       % professional-quality tables
\usepackage{amsfonts}       % blackboard math symbols
\usepackage{nicefrac}       % compact symbols for 1/2, etc.
\usepackage{microtype}      % microtypography
\usepackage{lipsum}
\usepackage{graphicx}
\graphicspath{ {./images/} }

\usepackage{svg}

\title{Utilizing Technical Data to Discover Similar Companies in Dhaka Stock Exchange}

\author{
 Tashreef Muhammad \\
  Department of Computer Science and Engineering \\
  Southeast University \\
  Dhaka, Bangladesh \\
  \texttt{tashreef.muhammad@seu.edu.bd} \\
   \And
 Tahsin Aziz \\
  Department of Computer Science and Engineering \\
  Ahsanullah University of Science and Technology \\
  Dhaka, Bangladesh \\
  \texttt{tahsinaziz.cse@aust.edu} \\
  \And
 Mohammad Shafiul Alam \\
  Department of Computer Science and Engineering \\
  Ahsanullah University of Science and Technology \\
  Dhaka, Bangladesh \\
  \texttt{shafiul.cse@aust.edu} \\
}

\begin{document}
\maketitle
\begin{abstract}
Stock market investment have been an ideal form of investment for many years. Investing capitals smartly in stock market yields high profit returns. But there are many companies available in a market. Currently there are more than $345$ active companies who have stocks in Dhaka Stock Exchange (DSE). Analyzing all these companies is quite impossible. However, many companies tend to move together. This study aims at finding which companies in DSE have a close connection and move alongside each other. By analyzing this relation, the investors and traders will be able to analyze a lot of companies' statistics from a calculating just a handful number of companies. The conducted experiment yielded promising results. It was found that though the system was not given anything other than technical data, it was able to identify companies that show domain specific outcomes. In other words, a relation between technical data and fundamental data was discovered from the conducted experiment. % From the data found through conducted experiments, a website is being build to help general public understand relation between different companies in an open-source environment and can be accessed using \href{https://github.com/TashreefMuhammad/CompanyGraph}{https://github.com/TashreefMuhammad/CompanyGraph}.
\end{abstract}

% keywords can be removed

\section{Introduction}
Stock markets provide an easy way for one to earn some fresh money. Through buying a stock of a company, the stockholder earns the title of owning a certain amount of that company. It makes the stockholder eligible to receive profit based on companies' income. It is actually quite similar to owning that business, only that the stockholder does not have to maintain all the tedious office work. However, it also mandates the stockholder to buy stock of companies that will gain profit, and more importantly do not get loss. To make such decisions, one needs to analyze different companies. There are couple of hundred of companies in different stock markets each and only in Dhaka Stock Exchange there are more than $345$ companies. Analyzing all companies is not a very feasible solution. But, it has been seen that a number of companies behave similarly. Hence, this study is an approach to try finding companies that tend to move together when it comes to price. \\

\subsection{Fundamental and Technical Data}
There are some specific understanding that help understand which companies usually move together. Usually, the data that helps identify such relation are the one's known as ``\textit{Fundamental Data}''. Fundamental data are those data that are related to the company itself, but is not portrayed as direct visible price that a stock contains on a specific moment. For example, it is known to many investors, that the ``\textit{Insurance}'' category companies tend to move together in DSE. But in the conducted study, our concentration was on ``\textit{Technical Data}'' which is the direct raw data visible for a stock in market. Ironically, through the conducted experiments, it was found that the known hypothesis from ``\textit{Fundamental Data}'' that the insurance companies move together, was actually supported through ``\textit{Technical Data}''. \\ 

\subsection{Motivation and Contribution}
Many studies have been conducted throughout the world on stock markets. However, on a scale to it, researches done on Dhaka Stock Exchange is very little. In addition, most of the conducted researches so far in the field is on predicting the closing price of stocks. For any investor or trader, knowing or predicting the price of stocks might be helpful, but other information also carry significant importance. Hence, this study pursues to find one such knowledge, that is to discover connection between companies through their technical data. The significant contribution of the study can be enlisted as:

\begin{itemize}
    \item Find correlation between DSE companies based on technical data only
    \item Discover evidence of technical data supporting hypothesis coming from fundamental data
    \item Find more specific real-world explainable connection between different companies
    \item Develop a visualization of the connection using graph to help visualize the system of companies in DSE
\end{itemize}

The rest of this paper is divided into some specific sections. Section ~\ref{sec:realted_work} discusses about studies on this field that have already been conducted. Section ~\ref{sec:experimental_setup} described how the data was collected and processed. It also describes how processed data was converted into a correlation matrix to find correlated companies. Then it discusses on how the visualization of the system was done using tools of graph theory. In Section ~\ref{sec:result_analysis}, the discussion is on the found results and some assumptions from the found results are discussed. Finally, in Section ~\ref{sec:conclusion} the conclusion is drawn with some references to future work.

\section{Related Works}
\label{sec:realted_work}

Most of the conducted researches in the field of stock markets are on predicting stock prices. Research related to stock price prediction using prediction techniques like neural networks has been ongoing for more than thirty years \cite{schoneburg1990stock}. 

Among several research works that have been conducted to predict stock price using Convolutional Neural Networks (CNNs) \cite{tsantekidis2017forecasting, gudelek2017deep, selvin2017stock, hiransha2018nse, kim2019forecasting, chen2018stock} have shown good performance. Since stock prices are time series data, they have property pf sequence data. Vanilla Recurrent Neural Networks (RNN) and Long-Short Term Memory (LSTM) models have been utilized \cite{selvin2017stock, hiransha2018nse, kim2019forecasting} for predicting stock prices as well.  Transformer based models for stock price prediction is also picking up pace. It has already been used to forecasting S\&P volatility \cite{ramos2021multi}. Transformer models have also been used on natural language data collected from social media related to stock price forecasting \cite{liu2019transformer}.

A number of researchers have used a variety of Artificial Intelligence (AI) techniques in stock price prediction \cite{obthong_survey_2020}. The so-called evolutionary and bio-inspired algorithms lead the deployment of meta-heuristics and AI-based techniques such as Genetic Algorithm, Artificial Bee Colony, Ant Colony, Fish Swarm optimization, Particle Swarm Optimization and the like \cite{SurveyBio}. Techniques of time series analysis like Box Jenkins method have also been used in some studies \cite{PSO_Box}.

This paper is on data from Dhaka Stock Exchange. Kamruzzaman et al. \cite{kamruzzaman2017modeling} published a study that uses Box-Jenkins methodology and applied Autoregressive Integrated Moving Average (ARIMA) to find interval forecasts of market return of DSE with 95\% confidence level. Maksuda et al. \cite{rubi2019forecasting} predicted the DSE Broad Index (DSEX) using a multi-layer feed-forward neural network and report satisfactory performance. Mujibur et al. \cite{BD_timeseriesforecast} deployed ARIMA, an artificial neural network, linear model, Holt-Winters model, and Holt-Winters exponential smoothing model on as many as 35 stocks of DSE and report the artificial neural network to perform relatively better compared to the other techniques. A recent study have also been conducted on DSE to predict stock prices using transformer based model \cite{muhammad2022transformer}. Alavi et al. \cite{alavi2021profitable} utilized different machine learning models for predicting the future using some factors. They also constructed a profit based ranking of different organizations based on observed accuracy and error rate.

There have been other applications of networks in stock market related works as well. Minjun Kim and Hiroki Sayama \cite{kim2017predicting} used network science for forecasting stock prices. Using correlation to analyze stock market network is not completely new. Quite the similar task was done by Wenyue Sun et al. \cite{sun2015network} in 2015 but with a completely different stock market based data and a completely different goal in mind. Piotr Szczepocki \cite{szczepocki2019clustering} used time-varying beta to study on Warsaw and Mansoor Momeni et al. \cite{momeni2015clustering} used k-means algorithm on Tehran Stock Exchange (TSE) to try and group similar companies. \\

A very recent survey on graph based works on stock markets were published by \cite{saha2022survey} that contains a very comprehensive collection of how graph-based approaches are being used in the field of stock markets. They have a completely dedicated section on discussion regarding using graphs for clustering companies.

The research gap that this study aims to overcome is developing a field for DSE companies to be classified in some clusters or groups. Stock data vary a lot from time to time, and also from place to place. The objective was to develop a very simple approach from which without much of a calculation a good analysis on companies of DSE can be found.

\section{Experimental Setup}
\label{sec:experimental_setup}

\subsection{Dataset Overview}
\label{subsec:dataset_overview}

For the purpose of this experiment, adjusted closing price of $386$ companies and three (3) market indices (00DS30, 00DSES, 00DSEX) were initially collected \cite{dataset}. The time line of collected data was from January 01, 2013 to July 13, 2022. The nature of collected data was End of Day (EoD) format and thus we had one row of data for each date for a specific company. Each row contained the date, opening price, highest price, lowest price, closing price and volume. For the experiment, only closing price was taken into consideration. \\
Later some pruning was done. Companies who did not have started at or before January 01, 2013 were removed. Similarly, companies that were closed before July 13, 2022 was moved out. Market indices are average of the market and they show relation between almost all the members. Hence, they were also removed. After all these pruning, there were $347$ companies left for the experiment to be conducted.

\subsection{Constructing Correlation Matrix}
\label{subsec:constructing_correlation_matrix}

\begin{figure}
    \centering
    \includegraphics[width = \textwidth, keepaspectratio]{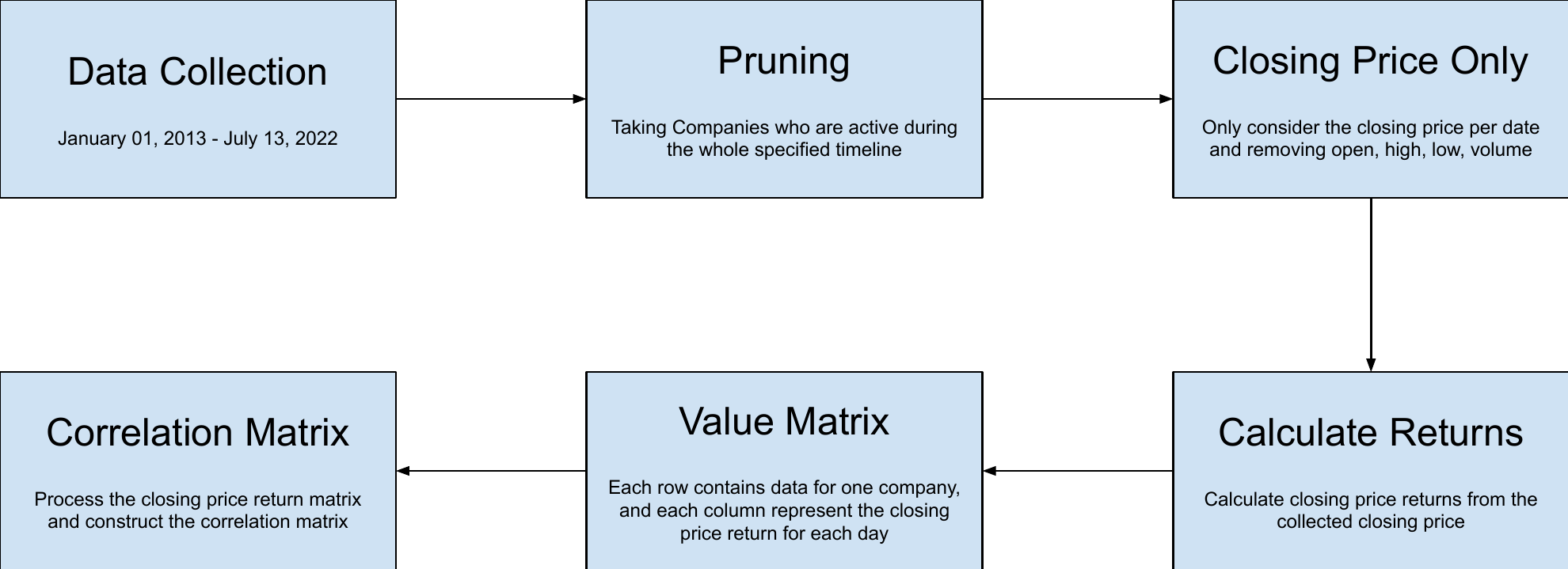}
    \caption{Construction of Correlation Matrix for the Experiment}
    \label{fig:construction_correlation}
\end{figure}

From the selected $347$ companies, first the closing price return was calculated. Closing price itself may vary a lot, but by calculating the difference, more knowledge can be gained about movement correspondence. Then a matrix was created where each column represented the closing price of each company for a certain date. After that, using all the data of different dates, the correlation matrix was created. Hence, the closing price return of each company from the timeline January 01, 2013 to July 13, 2022 was used to construct the correlation matrix. It was later on used for finding closely related companies. The whole process can be seen expressed as a diagram in Figure ~\ref{fig:construction_correlation}.

Pearson correlation coefficient formula was used for calculating correlation between two companies. Let us consider two companies are expressed using $X$ and $Y$. The closing price return on $i^{th}$ day for $X$ company expressed as $X_i$. Similarly, for $i^{th}$ day $Y_i$ represents closing price return of that day. $i$ represents the different dates in the data and $i \in [1, n]$. Then, the Pearson correlation can be expressed as,

\begin{equation}
    \rho_{X,Y} = \frac{cov(X, Y)}{\sigma_X \sigma_Y}
\end{equation}

where,
\begin{itemize}
    \item $cov(X, Y) = \frac{\sum_{i = 1}^n (X_i - \bar{X})(Y_i - \bar{Y})}{n}$
    \item $\sigma_X = $ Standard Deviation of $X$ 
    \item $\sigma_Y = $ Standard Deviation of $Y$
\end{itemize}
Using these tools the correlation between different companies were calculated and expressed through the correlation matrix. The values of Pearson correlation co-efficient lies between $[-1.0, 1.0]$ where,

\begin{itemize}
    \item $-1 = $ Complete Negative Correlation
    \item $ 0 = $ No Correlation
    \item $ 1 = $ Complete Positive Correlation
\end{itemize}

\subsection{Constructing Network}
\label{subsec:constructing_network}

\begin{figure}
    \centering
    \includegraphics[width=\textwidth]{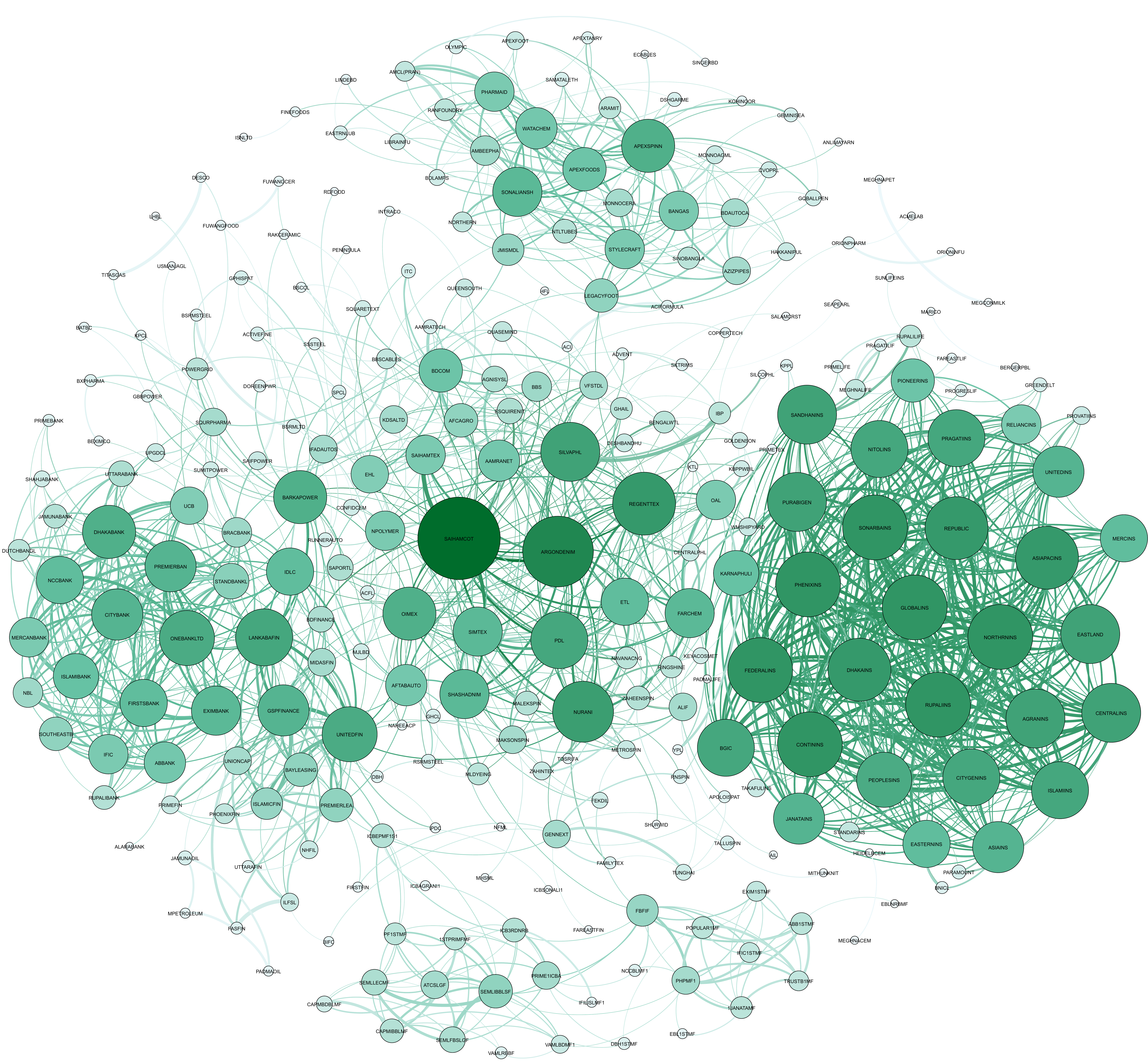}
    \caption{Network of Companies based on their closing price correlation}
    \label{fig:network}
\end{figure}

The objective of this study is to find companies that are interrelated to each other through the value found correlation matrix. It can be considered like formulating a network between different companies, where each company is connected by the value of their correlation coefficient. The stronger the correlation, the stronger is the connection between them. Now if it is considered that each company is a node and the correlation between them, is an edge that connects them, a graph can be constructed. \\
For the purpose of simplification we only took edge values of positive correlation and values greater than $0.5$. In that case the number of companies was reduced to $278$ and the number of found edges was $1393$. Using these data a network was constructed through Gephi \cite{bastian2009gephi} and it can be seen in Figure ~\ref{fig:network}.

It can be seen easily from Figure ~\ref{fig:network} that not all the companies in DSE have similar connections. In fact there are some visible cliques in the network. It can easily be assumed from the network that the companies forming close clusters tend to move quite alike. The found result through correlation matrix is quite large and hence this visual representation in Figure ~\ref{fig:network} provides a very concise and clear understanding of the findings of this study.

\section{Result Analysis}
\label{sec:result_analysis}

The performance of this study cannot be quantified using any particular metrics. The application of this study is purely evaluated through real-life intuition of the findings. Hence, some specific findings are highlighted in this section to verify the findings of the study with real-life scenarios.

\subsection{APEXFOODS, APEXSPINN, APEXFOOT are close}

In general APEXFOODS, APEXSPINN and APEXFOOT are from different sectors, But it can be seen that they are showing good interrelation with each other. It is mainly because they are part of same mother company. Though the fact that they are all part of the same company was not present in the data, still it was possible to get the interrelation between them.

\subsection{Insurance Companies are Closely Connected}

It is a common knowledge to traders of DSE that insurance companies tend to be very much interrelated. But during the process, there were no given data about which companies were insurance types. However, the found result easily shows that insurance companies are very much tightly clustered with each other.

\subsection{Bank and Financial Institutions}

It can be seen that banks and financial institutions are very close to each other. Also, from Figure ~\ref{fig:network} it can be seen that banks have more active relation with the whole network than financial institutions. The banks can also be seen to be connected to many distant companies in the network as well.

\subsection{Mutual Funds are Separated}

The mutual funds also show some level of difference from the rest of the networks and closeness among themselves. The graph in Figure ~\ref{fig:network} also shows that there can be made two parts of even among the mutual funds.

\subsection{Model Error Difference Calculation}

\begin{figure}
    \centering
    \includegraphics[width=.75\textwidth, keepaspectratio]{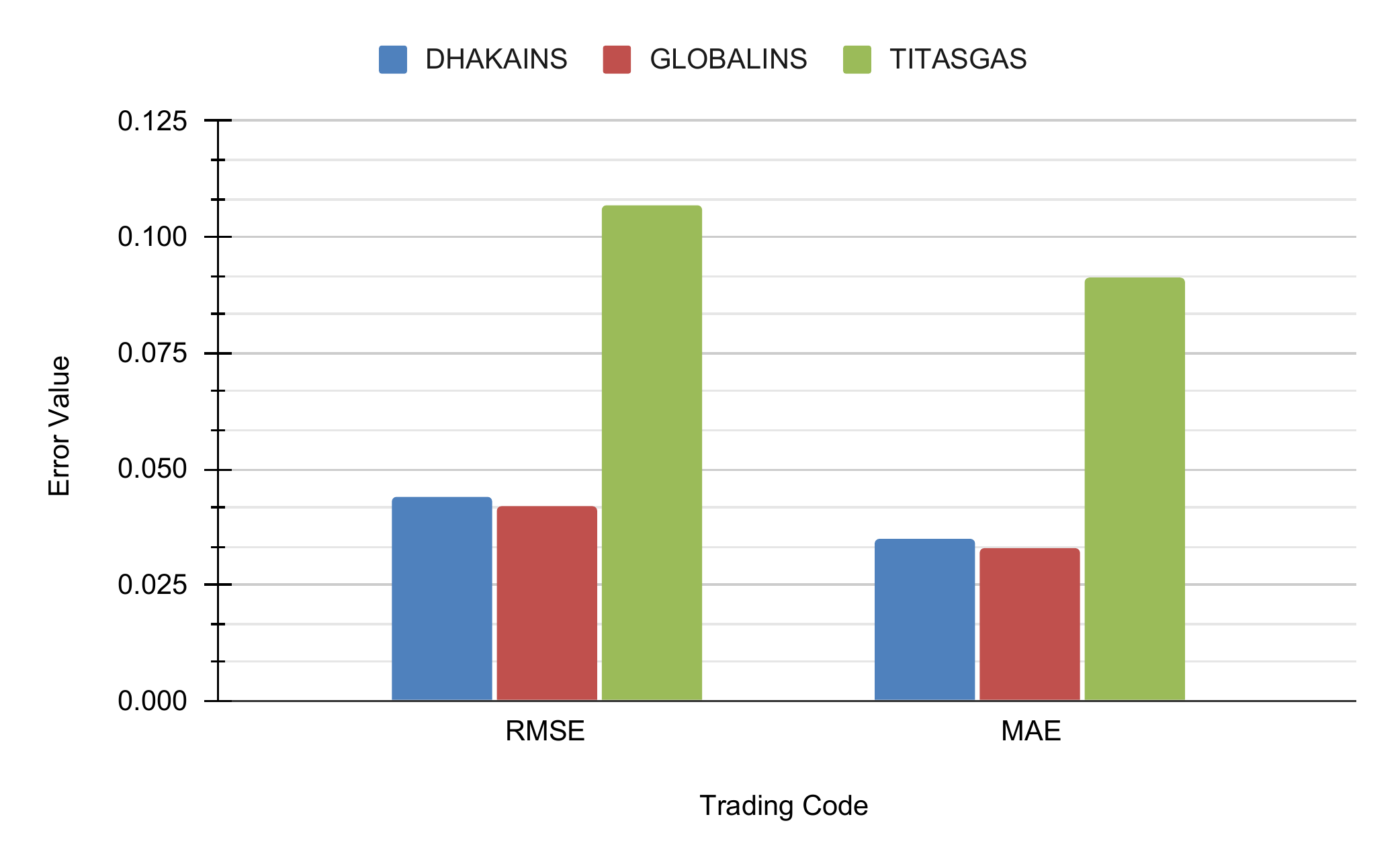}
    \caption{Forecasting Error Values for Three Different Companies}
    \label{fig:forecast_error}
\end{figure}

From the Figure ~\ref{fig:network} it can be seen that GLOBALINS is closely related to DHAKAINS but has a very far relation with TITASGAS. Using the model proposed by Tashreef et al. \cite{muhammad2022transformer} first we train a model using GLOBALINS and then apply that model on DHKAINS and TITASGAS. Because of the relation that is seen, TITASGAS prediction error should be much higher in contrast to DHAKINS prediction error. \\
After conducting the experiment, it was seen that truly the error for DHAKAINS and GLOBALINS was almost identical and very close, though GLOBALINS does have relatively smaller error. After all, the model was trained using GLOBALINS. But it still performed almost as well for DHAKAINS. However, the error value was much higher when the same model tried to forecast closing price for TITASGAS. The calculated Root Mean Squared Error (RMSE) and Mean Absolute Error (MAE) values for the three companies DHAKAINS, GLOBALINS and TITASGAS can be seen in Table ~\ref{tab:forecast_error}. From Figure ~\ref{fig:forecast_error} the relation between the forecasting error is very prominent.

\begin{table}[htbp]
    \centering
    \caption{Forecasting Error After Applying Machine Learning Model}
    \label{tab:forecast_error}
    \begin{tabular}{rcc}
    \hline
    \textbf{Trading Code}    &   \textbf{RMSE}  &   \textbf{MAE} \\ 
    \hline
    DHAKAINS	&   4.41E-02	&   3.48E-02    \\
    GLOBALINS	&   4.19E-02	&   3.29E-02    \\
    TITASGAS	&   1.07E-01	&   9.14E-02    \\
    \hline
    \end{tabular}
\end{table}

\section{Conclusion}
\label{sec:conclusion}

The aspects of this study is endless. Through analyzing domain data it can be made conclusive that the results that was found were quite accurate. Next, these values can be utilized by the investors to make better decision on which companies are similar that in the long run can help them invest in DSE. Also, the graph and data analysis might help find more connections present in Dhaka Stock Exchange that are yet not very possible to detect among the scattered fundamental data. Further study of DSE based on domain data will be greatly influenced by the study that has been conducted here. The most mention-able contribution of this study is the formation of a system that allots technical data from the market to present the fundamental relation of different companies. Being much more easily organized than the fundamental data, it will create many options for analysis of the Dhaka Stock Exchange, thus stock markets in total.

\bibliographystyle{unsrt}  
\bibliography{arXiv}

\end{document}